# Non-dipole effects in time delay of photoelectrons from atoms, negative ions, and endohedrals

M.Ya. Amusia[1, 2] and L.V. Chernysheva[2]

[1]*Racah Institute of Physics, the Hebrew University, Jerusalem, 91904 Israel*
[2]*Ioffe Physical-Technical Institute, St. Petersburg, 194021 Russia*

**Abstract**

In this Letter, we investigate the non-dipole effects in time delay of photoelectrons emitted by multi-electron atoms, negative ions, and respective endohedrals. We present the necessary general formulas in the frame of the random phase approximation with exchange (RPAE) applied to atoms, negative ions, and properly adjusted to endohedrals. We concentrate on low photon energy region, where non-dipole effects are very small in the cross-sections but become observable in angular distributions.

We not only derive the formulas for non-dipole effects in time delay, but perform corresponding numeric calculations. We demonstrate how the non-dipole corrections can be isolated in experiment. Concrete calculations are performed for noble gas atoms Ar and Xe, isoelectronic to them negative ions $Cl^-$ and I- and endohedrals $Ar(Cl^-)C_{60}$ and $Xe(I^-)@C_{60.}$ We found that the forward-backward photoelectron time delay differences give direct information on non-dipole effects. They proved to be quite measurable and prominently affected by the presence of the fullerenes shell.

**1.** The aim of this Letter is to find the contribution of non-dipole corrections to the time-delay of electrons that due to low-energy photon absorption leave atoms A, negative ions $A^-$ and endohedrals $A@C_N$ or $A^-@C_N$ that present A or $A^-$ stuffed inside the fullerenes shell $C_N$ constructed of N carbon atoms C. Ionization by a photon with low energy $\omega$[1] is a process determined predominantly by dipole transition matrix elements. The non-dipole contribution to the absolute photoionization cross-section is much smaller than the dipole term, being different by additional factor $(\omega r_A / c)^2 \ll 1$, where $r_A$ the atomic radius, and c is the speed of light. The smallness of the non-dipole cross-section is enhanced by a numeric small factor.

However, the role of non-dipole term is much bigger in the angular distributions of photoelectrons that differ from the respective dipole terms by a factor $(\omega r_A / c)$, which is much bigger than $(\omega r_A / c)^2 \ll 1$ (see e.g. [1]). The investigation of non-dipole transitions is stimulated by the interest to quadrupole transitions and quadrupole phase shifts of the outgoing electron wave-functions. Apart from angular distribution, there exists another source of information on these quantities, namely the time delay of electrons released from an atom, ion, or endohedral under the action of the incoming photon.

The theoretic approach to temporal description of processes in quantum objects has been developed long ago in a number of publications [2-5]. The construction of lasers with attosecond pulses permitted to measure the duration of atomic photoionization processes. As a result, pure theoretical development was relatively recently complemented by intensive experimental activity leading to temporal description of photoionization processes (see [6-10] and references therein). The corresponding times are called EWS time delays, thus referring to the names of authors L. Eisenbud [2], E. Wigner [4] and F. Smith [5], respectively.

[1] Atomic system of units $e = m = \hbar = 1$ is used in this paper, where $e$ is the electron charge, $m$ is its mass and $\hbar$ is Planck's constant.

As is demonstrated in [2-5], the time delays are well-defined quantities that characterize a physical process only if the outgoing particles interact via short range force. In course of photoionization the outgoing electron interacts with the residual ion with infinite range Coulomb interaction, making it impossible to define accurately enough scattering phases at any outgoing electron energy $\varepsilon > 0$. This forced to calculate time delays in photoionization starting from not too low $\varepsilon$ [7, 11]. This is why in this Letter we consider along with neutral targets, A and A@$C_N$, also negative ions that have the same electronic configuration as A, $A^-$ and $A^-$@$C_N$, for which the residual ion is neutral. In concrete calculations our choice are pairs Ar, $Cl^-$ and Xe, $I^-$. As $C_N$ we take an almost spherically symmetric well-studied and reasonably good modelled $C_{60}$ [12].

The time delay of a photoelectron emitted at a given angle at low $\omega$ is determined by the interference of electron waves emitted in a given direction due to absorption of a photon that include dipole and quadrupole terms. We will take into account the electron correlation in A($A^-$) and the action of $C_{60}$ upon A($A^-$) as well as the action of $C_{60}$ upon the outgaining electron. This will be done in the frame of the random phase approximation with exchange (RPAE), just as in [13, 14], substituting the $C_{60}$ by a static potential and adding to it the account of the polarization of $C_{60}$ electron shell by the incoming photon beam.

**2.** The temporal description of photoionization concentrates on determining the time delay of an electron that leaves the target in a given direction after absorbing a photon. As it was demonstrated in [2-5], the time delay $\tau$ of a physical process as a function of its energy $\varepsilon$ is connected to the phase of the amplitude $f$ of the process under consideration by the following relation

$$\tau(\varepsilon) = \frac{d}{d\varepsilon}\arg f(\varepsilon) = \frac{d}{d\varepsilon}\arctan\frac{\operatorname{Im} f(\varepsilon)}{\operatorname{Re} f(\varepsilon)} \equiv \operatorname{Im}\left[\frac{1}{f(\varepsilon)}\frac{df(\varepsilon)}{d\varepsilon}\right] \qquad (1)$$

The operator of photon-electron interaction with lowest order non-dipole corrections is presented in the so-called "length" form by the following expression [14]

$$\hat{M}_\kappa^r = \hat{d}_\kappa^r + \hat{q}_\kappa^\nabla \equiv \omega\left[(\mathbf{er}) + i(\boldsymbol{\kappa}\mathbf{r})(\mathbf{er})\right]. \qquad (2)$$

Here $\mathbf{e}$ is the photon polarization vector and $\boldsymbol{\kappa}$ is the photon momentum, $|\boldsymbol{\kappa}| = \omega / c$. For linearly polarized light one has

$$\hat{M}_\kappa^r = r\left[P_1(\cos\theta) + i\frac{\omega}{c} r P_2(\cos\theta)\right]. \qquad (3)$$

In the one-electron Hartree-Fock (HF) approximation the photoionization amplitude $f^{HF}(\mathbf{k})$ is determined by the following matrix element $\left\langle \psi_{\mathbf{k}}^{(-)} \middle| \hat{M}_\kappa^r \middle| \phi_i \right\rangle$, which describes the transition of an atomic electron from initial state $i$ with the wave-function $\phi_i(\mathbf{r}) = R_{n_i l_i}(r) Y_{l_i m_i}(\hat{\mathbf{r}})$ to the continuous state of an electron with momentum $\mathbf{k}$ after absorbing a linearly polarized photon, in which case the polarization vector is directed along the z axis and $z = \sqrt{4\pi/3} r Y_{10}(\hat{\mathbf{r}})$ [2] [15, 7]. Thus, in HF approximation $f(\omega) \Rightarrow f_{n_i l_i}^{HF}(\varepsilon, \hat{\mathbf{k}})$ is determined by the following formula that is similar to (2) in [11], but with another reduced radial matrix element.

$$
\begin{array}{l}
f_{n_i l_i}^{HF}(\varepsilon, \hat{\mathbf{k}}) = \dfrac{(2\pi)^{3/2}}{k^{1/2}} \left[ \sum_{l = l_i \pm 1} e^{i[\delta_l(\varepsilon) - l\pi/2]} Y_{lm_i}(\hat{\mathbf{k}}) \begin{pmatrix} l & 1 & l_i \\ -m_i & 0 & m_i \end{pmatrix} \langle \varepsilon l \| r \| n_i l_i \rangle + \right. \\
\left. \dfrac{\omega}{c} \sum_{l' = l_i \pm 2, 0} e^{i[\delta_{l'}(\varepsilon) - l'\pi/2 + \pi/2]} Y_{l'm}(\hat{\mathbf{k}}) \begin{pmatrix} l' & 2 & l_i \\ -m_i & 0 & m_i \end{pmatrix} \langle \varepsilon l' \| r^2 \| n_i l_i \rangle \right]
\end{array}. \quad (4)
$$

Here $\varepsilon = k^2/2$, $l, l'$ denotes the angular momentum of the outgoing photoelectron and $n_i l_i$ are the principal quantum number and angular momentum of the ionized electron. The continuous spectrum radial wave function are normalized in the energy scale according to the relation $\langle \varepsilon l | \varepsilon' l \rangle = \delta(\varepsilon - \varepsilon')$ and their asymptotic is given by the following formula [16]

$$
R_{\varepsilon l}(r)\big|_{r \to \infty} = \sqrt{\frac{2}{\pi k}} \frac{1}{r} \sin\left( kr - \frac{l\pi}{2} + \delta_l \right). \quad (5)
$$

Here $\delta_l \equiv \delta_l(\varepsilon)$ is the continuous spectrum electron scattering phase in HF approximation [13-15].

Usually, in attosecond time delay measurements the forward direction amplitude is of interest [7]. In this case one has to substitute $Y_{lm}(\hat{\mathbf{k}} = 0)$ by $\sqrt{(2l+1)/4\pi}\delta_{m0}$. In current Letter, in order to better separate the non-dipole contribution from dipole, we will need also the backward direction amplitudes that can be obtained from (4) substituting there $Y_{lm}(\hat{\mathbf{k}} = \pi)$ by $(-1)^l \sqrt{(2l+1)/4\pi}\delta_{m0}$. In order to investigate also angular dependence of the time delay at any $\theta$ in the simplest way, let us study only the case $m_i = 0$, so that $Y_{lm_i}(\theta_{\mathbf{k}}, \varphi_{\mathbf{k}})$ has to be substituted by $\sqrt{(2l+1)/4\pi} P_l(\cos\theta)$. As is seen from (1), time delay is not affected by real numeric factors in the total amplitude, since they do not affect the phase shift. Therefore we can define the photoionization amplitude neglecting common factors. So, for the aims of time delay determination we have in HF, omitting factor $2^{1/2}\pi/k^{1/2}$, instead of (4) the following relation

$$
\begin{array}{l}
f_{n_i l_i}^{HF}(\varepsilon, \theta_{\mathbf{k}}) \equiv f_{n_i l_i, d}^{HF}(\varepsilon, \theta_{\mathbf{k}}) + f_{n_i l_i, q}^{HF}(\varepsilon, \theta_{\mathbf{k}}) = \left[ \sum_{l = l_i \pm 1} e^{i[\delta_l(\varepsilon) - l\pi/2]} P_l(\cos\theta_{\mathbf{k}}) \sqrt{2l+1} \begin{pmatrix} l & 1 & l_i \\ 0 & 0 & 0 \end{pmatrix} \times \right. \\
\left. \langle \varepsilon l \| r \| n_i l_i \rangle + \dfrac{\omega}{c} \sum_{l' = l_i \pm 2, 0} e^{i[\delta_{l'}(\varepsilon) - l'\pi/2 + \pi/2]} P_{l'}(\cos\theta_{\mathbf{k}}) \sqrt{2l'+1} \begin{pmatrix} l' & 2 & l_i \\ 0 & 0 & 0 \end{pmatrix} \langle \varepsilon l' \| r^2 \| n_i l_i \rangle \right]
\end{array}. \quad (6)
$$

---

[2] A roof sign above a vector as e.g. in $\hat{\mathbf{r}}$ means a unit vector in the direction of vector $\mathbf{r}$.

Here $\varepsilon = \omega - \varepsilon_i = k^2/2$, while $\delta_l(\varepsilon)$, $\delta_{l'}(\varepsilon)$ are photoelectron phases in HF approximation. It is seen that at a given angle the dipole (first term) and quadrupole (the second term) amplitudes in (6) interfere leading to different time delays at different angles $\theta$.

To obtain the photoionization amplitudes that take into account the interelectron correlation in RPAE, one has to substitute in (6) the matrix elements $\langle \varepsilon l \| r \| n_i l_i \rangle$ and $\langle \varepsilon l' \| r^2 \| n_i l_i \rangle$ by $\langle \varepsilon l \| D(\omega) \| n_i l_i \rangle$ and $\langle \varepsilon l' \| Q(\omega) \| n_i l_i \rangle$ that are determined by the following equations [13, 14]:

$$\begin{gathered}\langle \varepsilon l \| D(\omega) \| n_i l_i \rangle = \langle \varepsilon l \| r \| n_i l_i \rangle + \frac{1}{3} \sum_{\substack{n'l' \\ n_j l_j}} \int_0^\infty d\varepsilon' \left[ \frac{\langle \varepsilon' l' \| D(\omega) \| n_j l_j \rangle \langle n_j l_j, \varepsilon l \| U \| \varepsilon' l', n_i l_i \rangle}{\omega - \varepsilon' + \varepsilon_{n_i l_i} + i\eta} \right. \\ \left. + \frac{\langle n_j l_j \| D(\omega) \| \varepsilon' l' \rangle \langle \varepsilon' l', \varepsilon l \| U \| n_j l_j, n_i l_i \rangle}{\omega + \varepsilon' - \varepsilon_{n_i l_i}} \right], \quad \eta \to +0\end{gathered}, \quad (7)$$

and

$$\begin{gathered}\langle \varepsilon l' \| Q(\omega) \| n_i l_i \rangle = \langle \varepsilon l' \| r^2 \| n_i l_i \rangle + \frac{1}{5} \sum_{\substack{n''l'' \\ n_j l_j}} \int_0^\infty d\varepsilon' \left[ \frac{\langle \varepsilon'' l'' \| Q(\omega) \| n_j l_j \rangle \langle n_j l_j, \varepsilon l' \| U \| \varepsilon'' l'', n_i l_i \rangle}{\omega - \varepsilon'' + \varepsilon_{n_i l_i} + i\eta} \right. \\ \left. + \frac{\langle n_j l_j \| Q(\omega) \| \varepsilon'' l'' \rangle \langle \varepsilon'' l'', \varepsilon l' \| U \| n_j l_j, n_i l_i \rangle}{\omega + \varepsilon'' - \varepsilon_{n_i l_i}} \right], \quad \eta \to +0\end{gathered}. \quad (8)$$

In (7) and (8) summation over $n_j l_j$ includes all occupied electron subshell of an atom, while summation over $n'l'(n''l'')$ includes along with discrete vacant excited states also integration over continuous spectrum one-electron excitations. Here matrix elements of $U$ are combinations of direct and exchange Coulomb interelectron interaction $V = 1/|\mathbf{r}_1 - \mathbf{r}_2|$. More details on solving (7, 8) and corresponding computing programs can be found in [14].

Thus, with the help of (7) and (8) the RPAE amplitude is given by the following relation:

$$\begin{gathered}f_{n_i l_i}^{RPAE}(\varepsilon, \theta_{\mathbf{k}}) \equiv f_{n_i l_i, d}^{RPAE}(\varepsilon, \theta_{\mathbf{k}}) + f_{n_i l_i, q}^{RPAE}(\varepsilon, \theta_{\mathbf{k}}) = \\ \left[ \sum_{l = l_i \pm 1} e^{i[\delta_l(\varepsilon) - l\pi/2]} P_l(\cos\theta_{\mathbf{k}}) \sqrt{2l+1} \begin{pmatrix} l & 1 & l_i \\ 0 & 0 & 0 \end{pmatrix} \langle \varepsilon l \| D(\omega) \| n_i l_i \rangle + \right. \\ \left. \frac{\omega}{c} \sum_{l' = l_i \pm 2, 0} e^{i[\delta_{l'}(\varepsilon) - l'\pi/2 + \pi/2]} P_{l'}(\cos\theta_{\mathbf{k}}) \sqrt{2l'+1} \begin{pmatrix} l' & 2 & l_i \\ 0 & 0 & 0 \end{pmatrix} \langle \varepsilon l' \| Q(\omega) \| n_i l_i \rangle \right]\end{gathered} \quad (9)$$

In calculating $\arg f_{n_i l_i}^{RPAE}(\varepsilon, \theta_{\mathbf{k}})$ we have to include contributions that comes from singularity in the first terms in square brackets in (7, 8), determined by $\eta \to +0$. Since $\mathrm{Re}\langle \varepsilon l \| D(\omega) \| n_i l_i \rangle$ and $\mathrm{Re}\langle \varepsilon l' \| Q(\omega) \| n_i l_i \rangle$ as functions of $\omega$ change sign thus going via zero, the phases $\arg f_{n_i l_i}^{RPAE}(\varepsilon, \theta_{\mathbf{k}})$

jump from $\pi/2$ to $-\pi/2$ at respective points in $\omega$ and $\varepsilon$. These jumps make numeric calculations considerably more complex. But at these points the derivative $d[\arg f(\varepsilon)]/d\varepsilon$ that enters the time delay definition (1) is a smooth function. So, instead of (1) we employ in RPAE an equivalent formula:

$$\tau_{n_i l_i}^{RPAE}(\varepsilon,\theta_{\mathbf{k}}) = \frac{\left[d\,\mathrm{Im} f_{n_i l_i}^{RPAE}(\varepsilon,\theta_{\mathbf{k}})/d\varepsilon\right]\mathrm{Re} f_{n_i l_i}^{RPAE}(\varepsilon,\theta_{\mathbf{k}}) - \left[d\,\mathrm{Re} f_{n_i l_i}^{RPAE}(\varepsilon,\theta_{\mathbf{k}})/d\varepsilon\right]\mathrm{Im} f_{n_i l_i}^{RPAE}(\varepsilon,\theta_{\mathbf{k}})}{\left[\mathrm{Re} f_{n_i l_i}^{RPAE}(\varepsilon,\theta_{\mathbf{k}})^2 + \mathrm{Im} f_{n_i l_i}^{RPAE}(\varepsilon,\theta_{\mathbf{k}})^2\right]}. \tag{10}$$

**2.** To investigate time-delays in endohedrals A@C$_N$, we substitute $C_{60}$ by a static potential *W*(*r*) that corresponds to reasonable distribution of electric charges and satisfactorily reproduces the experimental value of an s-electron affinity of $C_{60}^-$, just as it was done recently in [17]. We call this approach RPAE$_C$. The potential *W*(*r*) is added to the HF atomic A potential. It affects HF$_C$ and RPAE$_C$ matrix elements and phases. The chosen $C_{60}$ potential is the same as in [18]

$$W(r) = -W_{\max} d^2/[(r-R)^2 + d^2]. \tag{11}$$

In addition to the static potential *W*(*r*) it is essential to take into account the polarization of the fullerene shell that essentially modifies the action of the incoming light beam upon the ionized inner atom A. Assuming for simplicity that the atomic radius $R_a$ is much smaller than *R*, one can in RPAE frame express the effect of dipole and quadrupole polarization by factors $G_{C_N}^d(\omega)$ and $G_{C_N}^q(\omega)$ that connects the endohedral photoionization amplitude in "atomic" RPAE $f_{n_i l_i}^{RPAE_C}(\varepsilon)$ with the endohedral RPAE photoionization amplitude $f_{n_i l_i}^{A@C_N}(\varepsilon)$ by a simple relation

$$f_{n_i l_i}^{A@C_N}(\varepsilon,\theta_{\mathbf{k}}) = G_{C_N}^d(\omega) f_{n_i l_i,d}^{RPAE_{C_N}}(\varepsilon,\theta_{\mathbf{k}}) + G_{C_N}^q(\omega) f_{n_i l_i,q}^{RPAE_{C_N}}(\varepsilon,\theta_{\mathbf{k}}). \tag{12}$$

The polarization factors $G_{C_N}^d(\omega)$ and $G_{C_N}^d(\omega)$ are expressed via dipole and quadrupole polarizabilities of $C_{60}$ [13, 14]:

$$G_{C_N}^d(\omega) \simeq \left[1-\alpha_C^d(\omega)/R^3\right],\ G_{C_N}^q(\omega) \simeq \left[1-\alpha_C^q(\omega)/4R^5\right]. \tag{13}$$

Here $\alpha_C^d(\omega)$ and $\alpha_C^q(\omega)$ are the $C_{60}$ dipole and quadrupole dynamic polarizabilities. The dipole polarizability is known quite well, and is determined by the total photoionization cross section $\sigma(\omega)$ of $C_N$ (see, e.g.[13]) that is measurable and limited by the so-called dipole sum-rule. As to $\alpha_C^q(\omega)$, there is no experimental data that could help to determine this quantity. Using Eq. (9), (12) and (13), we obtain the following expression for the endohedral amplitude with non-dipole corrections $f_{n_i l_i}^{A@C_N}(\varepsilon,\theta_{\mathbf{k}})$:

$$f_{n_il_i}^{A@C_N}(\varepsilon,\theta_{\mathbf{k}}) \equiv f_{n_il_i,d}^{A@C_N}(\varepsilon,\theta_{\mathbf{k}}) + f_{n_il_i,q}^{A@C_N}(\varepsilon,\theta_{\mathbf{k}}) =$$

$$\left[G_{C_N}^d(\omega)\sum_{l=l_i\pm1} e^{i[\delta_l^C(\varepsilon)-l\pi/2]}P_l(\cos\theta_{\mathbf{k}})\sqrt{2l+1}\begin{pmatrix} l & 1 & l_i \\ 0 & 0 & 0\end{pmatrix}\langle\varepsilon l\|D(\omega)\|n_il_i\rangle + \right. \tag{14}$$

$$\left.\frac{\omega}{c}G_{C_N}^d(\omega)\sum_{l'=l_i\pm2,0} e^{i[\delta_{l'}^C(\varepsilon)-l'\pi/2+\pi/2]}P_{l'}(\cos\theta_{\mathbf{k}})\sqrt{2l'+1}\begin{pmatrix} l' & 2 & l_i \\ 0 & 0 & 0\end{pmatrix}\langle\varepsilon l'\|Q(\omega)\|n_il_i\rangle\right]$$

Note that phases $\delta_l^C(\varepsilon)$ are determined in $HF_C$ approximation. The time delays for endohedrals $\tau_{n_il_i}^{A@C_N}(\varepsilon,\theta_{\mathbf{k}})$ are given by a formula similar to (10) where $f_{n_il_i}^{RPAE}(\varepsilon,\theta_{\mathbf{k}})$ is substituted by $f_{n_il_i}^{A@C_N}(\varepsilon,\theta_{\mathbf{k}})$.

In low-energy photoionization studies the parameter that characterizes the contribution of the quadrupole term as compared to the dipole one $\omega r_A/c$ is small, even at $(\omega r_A/c)\ll 1$. However, the non-dipole terms are observed relatively easy at Cooper or interference minima in the dipole amplitude [13], or in cases when due to e.g. resonance phenomena the quadrupole terms are unusually big.

For subvalent subshells *ns* of noble gas atoms the dipole matrix elements are abnormally suppressed while the quadrupole terms are enhanced [13]. The same is valid to large extent for subvalent subshells in noble gas endohedrals. This is why s-subshells are of special interest. For s- ionized electrons, i.e. for $l_i = 0$, one has from (9) and (14), omitting a common in both terms an inessential for time delay factor *i*, the following formulas

$$\begin{aligned} f_{n_i0}^{RPAE}(\varepsilon,\theta_{\mathbf{k}}) &= e^{i\tilde{\delta}_1}\cos\theta_{\mathbf{k}}\tilde{D}_1 - \sqrt{2}\frac{\omega}{c}e^{i\tilde{\delta}_2}P_2(\cos\theta_{\mathbf{k}})\tilde{Q}_2, \\ f_{n_i0}^{A@C_N}(\varepsilon,\theta_{\mathbf{k}}) &= G^d e^{i\tilde{\delta}_1}\cos\theta_{\mathbf{k}}\tilde{D}_1 - \sqrt{2}\frac{\omega}{c}G^q e^{i\tilde{\delta}_2(\varepsilon)}P_2(\cos\theta_{\mathbf{k}})\tilde{Q}_2 \end{aligned}. \tag{15}$$

Here $\langle\varepsilon 1\|D(\omega)\|n_i0_i\rangle \equiv \tilde{D}_1\exp[i\Delta_1(\varepsilon)]$, $\tilde{\delta}_1 \equiv \delta_1(\varepsilon)+\Delta(\varepsilon)$, $\langle\varepsilon 2\|Q(\omega)\|n_i0_i\rangle \equiv \tilde{Q}_2\exp[i\Delta_2(\varepsilon)]$, and $\tilde{\delta}_2 \equiv \delta_2(\varepsilon)+\Delta_2(\varepsilon)$.

The time delay for s-subshells in RPAE is particularly simple for $\omega/c \ll \cos\theta_{\mathbf{k}}$ where it is given by the formula:

$$\tau_{n_i0}^{RPAE}{}_2(\varepsilon,\theta_{\mathbf{k}}) \approx \tilde{\delta}_1' - \frac{\sqrt{2}}{c}\frac{P_2(\cos\theta_{\mathbf{k}})}{\cos\theta_{\mathbf{k}}}\left[\left[\frac{\tilde{Q}_2}{\tilde{D}_1}+\omega\left(\frac{\tilde{Q}_2}{\tilde{D}_1}\right)'\right]\sin(\tilde{\delta}_2-\tilde{\delta}_1)+\omega\frac{\tilde{Q}_2}{\tilde{D}_1}(\tilde{\delta}_2'-\tilde{\delta}_1')\cos(\tilde{\delta}_2-\tilde{\delta}_1)\right]. \tag{16}$$

To compare the non-dipole terms for $\tau_{n_i0}^{RPAE}(\varepsilon,\theta_{\mathbf{k}})$ with that for the cross-section $\sigma_{n0}(\omega)$ let us present the corresponding formula [13]:

$$\frac{d\sigma_{n0}(\omega,\theta_{\mathbf{k}})}{d\Omega} \approx \frac{3\sigma_{n0}(\omega)}{8\pi}\sin^2\theta_{\mathbf{k}}[1+4\frac{\omega}{c}\frac{\tilde{Q}_2}{\tilde{D}_1}\cos(\tilde{\delta}_2-\tilde{\delta}_1)\cos\theta_{\mathbf{k}}]. \qquad (17)$$

To obtain expressions similar to (15)-(17) but for endohedrals, one has to substitute $\tilde{Q}_2$, $\tilde{D}_1$, $\tilde{\delta}_2$, and $\tilde{\delta}_1$ by $\tilde{\tilde{Q}}_2 = \tilde{G}^q\tilde{Q}_2$, $\tilde{\tilde{D}}_1 = \tilde{G}^d\tilde{D}_1$, $\tilde{\tilde{\delta}}_2^C = \tilde{\delta}_2^C + \Delta^q$, and $\tilde{\tilde{\delta}}_1^C = \tilde{\delta}_1^C + \Delta^p$, where the following notations $\tilde{G}^{d,q}\exp(i\Delta^{d,q}) \equiv G_{C_N}^{d,q}(\omega)$ are employed.

According to (15), by measuring the time-delay at $\theta_{\mathbf{k}} = \pi/2$ one can find directly the non-dipole contribution since the dipole term in (15) is equal to zero. However, according to (17) the this angle is zero. It is essential that the quadrupole term enters time-delay and photoionization cross-section in different combinations.

It is essential to note that by going from observation angle $\theta_{\mathbf{k}}$ to $\theta_{\mathbf{k}}+\pi$, from forward to backward directions, for example, the dipole contribution does not change its sign, while the quadrupole contribution changes its signs. Thus, by measuring the differences $\Delta\tau_{n_il_i}^{RPAE}(\varepsilon,\theta_{\mathbf{k}}) \equiv \tau_{n_il_i}^{RPAE}(\varepsilon,\theta_{\mathbf{k}}) - \tau_{n_il_i}^{RPAE}(\varepsilon,\theta_{\mathbf{k}}+\pi)$ or $\Delta\tau_{n_il_i}^{A@C_N}(\varepsilon,\theta_{\mathbf{k}}) \equiv \tau_{n_il_i}^{A@C_N}(\varepsilon,\theta_{\mathbf{k}}) - \tau_{n_il_i}^{A@C_N}(\varepsilon,\theta_{\mathbf{k}}+\pi)$ one can obtain quantities that are directly proportional to $\omega/c$.

The concrete analytic expressions for time differences $l_i > 0$ are too cumbersome to be exposed here. Instead, we will present directly the numeric results of calculations of time differences. Note that the difference between $d\sigma_{n_il_i}(\varepsilon,\theta_{\mathbf{k}})/d\Omega$ at $\theta_{\mathbf{k}}$ and $\theta_{\mathbf{k}}+\pi$ is also proportional to $1/c$ that is illustrated for s-subshell by (17).

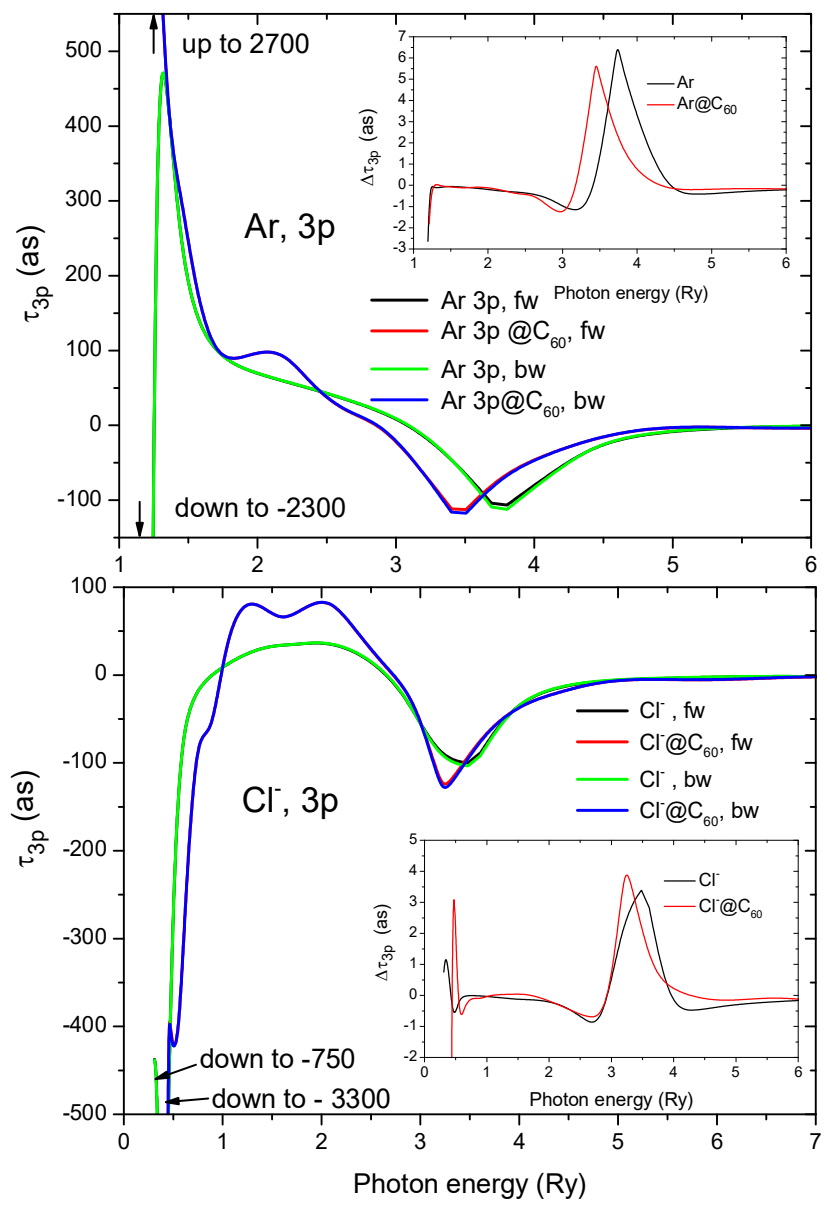

Fig. 1. Time delays of photoelectrons from 3p of Ar, $Cl^-$ and endohedrals $Ar@C_{60}$, $Cl^-@C_{60}$, emitted forward (fw) and backward (bw) and their differences $\Delta\tau_{3p}(\varepsilon) \equiv \tau_{3p}(\varepsilon,0) - \tau_{3p}(\varepsilon,\pi)$.

It is seen from (9), (14), (15), and (17) that the quadrupole term adds new information about the photoionization process: includes another photoionization amplitude and polarization factor, other scattering phase differences. It is essential that these characteristics appear in combinations different from that in non-dipole corrections to the angular distribution of photoelectrons (see (17)) [13]. Therefore the results that could be obtained from studies of the non-dipole corrections to the time delay are complimentary to that obtained from the angular distributions.

**3.** We have performed concrete calculations for multi-electron atoms Ar, Xe, isoelectronic to them negative ions $Cl^-$, $I^-$ and respective endohedrals, consisting of $C_{60}$ fullerene shell with one of atoms/ions inside: Ar, $Cl^-$, Xe, $I^-@C_{60}$. We considered outer p- and subvalent s-subshells in all studied objects along with 4d in Xe and $I^-$, as well as in their endohedrals. We concentrate here on obtaining data on non-dipole contributions. Calculations are performed in $RPAE_C$. The parameters of the potential (11) are the same as in our recent papers, e.g. in [11]. Inclusion of fullerenes shell polarization in the RPAE frame is achieved by using G-factors (13), the parameters of which are the same as in [11]. The results for endohedrals are marked by upper indexes A@C.

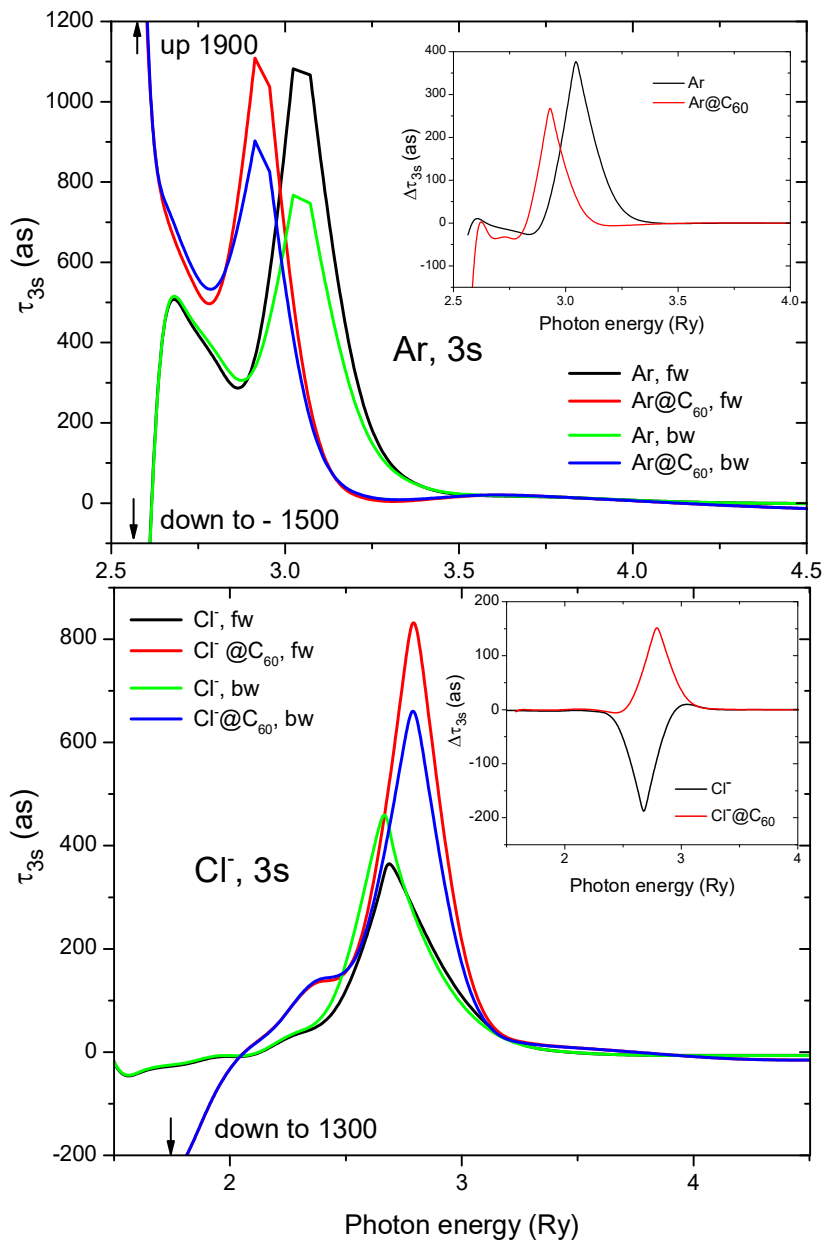

Fig. 2. Time delays of photoelectrons from 3s of Ar, Cl- and endohedrals Ar@$C_{60}$, $Cl^-$@$C_{60}$, emitted forward (fw) and backward (bw) and their differences $\Delta\tau_{3s}(\varepsilon) \equiv \tau_{3s}(\varepsilon,0) - \tau_{3s}(\varepsilon,\pi)$.

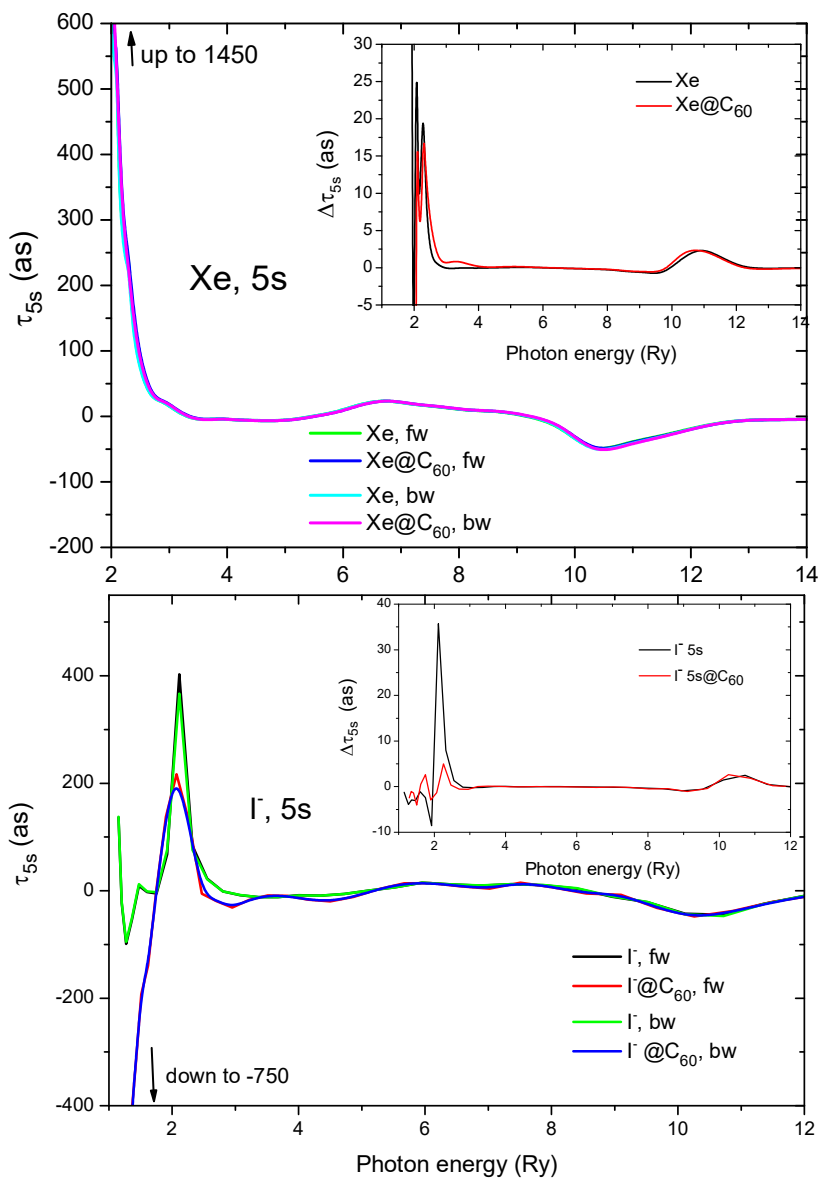

Fig. 3. Time delays of photoelectrons from 5s of Xe, I- and endohedrals Xe@$C_{60}$, $I^-$@$C_{60}$, emitted forward (fw) and backward (bw) and their differences $\Delta\tau_{5s}(\varepsilon) \equiv \tau_{5s}(\varepsilon,0) - \tau_{5s}(\varepsilon,\pi)$.

In calculations we determine only HF scattering phases $\delta_l(\varepsilon)$ (9) and (14). These phases are well defined only for short-range potentials. Such is the case for photoionization of negative ions and respective endohedrals. For neutral atoms and their endohedrals the outgoing electron feels the single charged ion's field. As a result, the very notion of a phase shift loses its sense. Trial calculations demonstrate that for considered objects it took place for $\varepsilon$<0.1 Ry. For negative ions the time-delay is well defined at any energy.

Using (10) we calculate the $RPAE_C$ and (after proper substitution of amplitudes) A@C forward and backward time delays at the photoelectron emission angle $\theta_{\mathbf{k}} = 0$ and $\theta_{\mathbf{k}} = \pi$, denoted in Fig. 1-4 as fw and bw, respectively. The main finding here are time differences that are proportional to $\omega / c$. They are presented as insets in these figures. The photon energy is in Rydbergs (Ry, 1Ry=13.6 eV) and time delay is in Attoseconds (as, 1as=$10^{-18}$sec). We are not presenting here results of calculations in HF, since it is known that for the considered objects the role of RPAE corrections is big [13].

Each figure includes for a given subshell results for an atom, isoelectronic negative ion and respective endohedrals. Big negative values of time delay contradict the causality principle and as such has to be discarded. However, not too big negative values are compatible with causality due to quantum nature of the photoionization process (see [4]). The forward and backward time delays are quite big reaching hundreds attoseconds. The respective forward-backward time difference is characterized in Ar by a strong maximum that is quite noticeable for 3p –subshell (see Fig. 1) and particularly big for 3s-subshell in Ar and Ar@$C_{60}$ (see Fig. 2). The influence of the endohedral shell upon time delays and corresponding non-dipole corrections is prominent, but only for $Cl^-$ and $Cl^-$

@$C_{60}$ a high maximum in $\Delta\tau_{3s}(\varepsilon)$ transforms into a deep minimum (see Fig. 2, inset). The strong variation in $\Delta\tau_{3s}(\varepsilon)$ is a consequence of smallness of dipole photoionization amplitudes and big influence upon them of quadrupole transition [1, 13]. The behavior of time delay for 3p electrons follows the dependence of 3p dipole amplitude upon $\omega$ and its complete domination over the contribution of quadrupole transition.

We do not present results for 5p in Xe and $I^-$ as well as respective endohedrals since the situation there is qualitatively similar to 3p in Fig. 1 and due to lack of space. Fig. 3 depicts the results for 5s-photoelectrons. 5s photoionization amplitude for considered objects is strongly affected not only by outer 5p electrons but by the intermediate 4d-subshell also. The well-known feature of 4d is its dipole Giant resonance that dominates the cross-section in a broad $\omega$ region – from about 4 to 10Ry. However, in 5s time delay important is only the maximum that reflects the 5p influence upon 5s. Surprisingly enough, only small traces of the Giant resonance exist. The maxima in insets of Fig. 3 reflect the influence of the fact that the dipole photoionization cross section at about 11 Ry has a Cooper minimum [13]. Fig. 4 demonstrates results for 4d subshell. As for 5p, the near threshold behavior is of interest, presenting quite different results for Xe and $I^-$ No essential traces of the Giant resonance is observed. However in the region of Cooper minimum in the photoionization cross-section we see a prominent minimum in time delays. The time differences for Xe and $I^-$ has a remarkable variation of Fano profile shape that does not correspond, however, to a discrete level. Instead, it is a trace of dipole cross-section variation. The polarization factor affects mainly the outer subshell, and leads to rather smooth addition to the time delay near thresholds of outer and subvalent subshells in Ar, $Cl^-$, Xe, $I^-$ and the respective endohedrals. Naturally, the 4d subshell is deep enough, so the role of $C_{60}$ shell is for all time delays negligible.

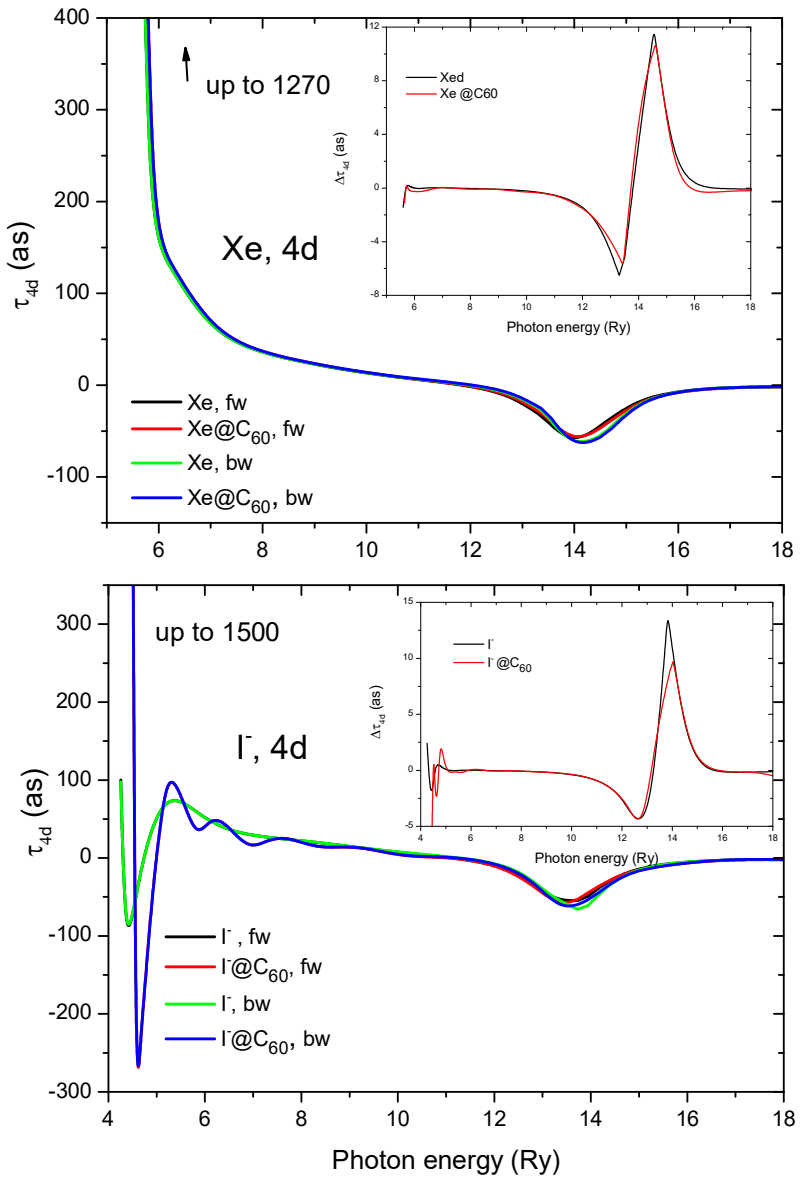

Fig. 4. Time delays of photoelectrons from 4d of Xe, $I^-$ and endohedrals Xe@$C_{60}$, $I^-$@$C_{60}$, emitted forward (fw) and backward (bw) and their differences $\Delta\tau_{4d}(\varepsilon) \equiv \tau_{4d}(\varepsilon,0) - \tau_{4d}(\varepsilon,\pi)$.

**4.** We investigated here the photoionization amplitude at low photon energies that includes along the main dipole also quadrupole contribution. Having this amplitude, we demonstrate for the first time, that quadrupole corrections are quite noticeable and under concrete conditions can be isolated from the dipole ones, and measured. Of particular interest is the situation for 3s subshell in Ar, where the determined by quadrupole contribution time difference reaches 400 *as*.

To use attosecond lasers in studies of time-delays is not a simple task since extraction from experiment the EWS time delays require to get rid of the additions introduced by the properties of the laser beam including the uncertainty in energy of the incoming photons (see e.g. [19, 11]).

In spite of difficulties, the theoretical investigation of time delay differences and their experimental measurement is of interest as a source of unique information about ionized targets.

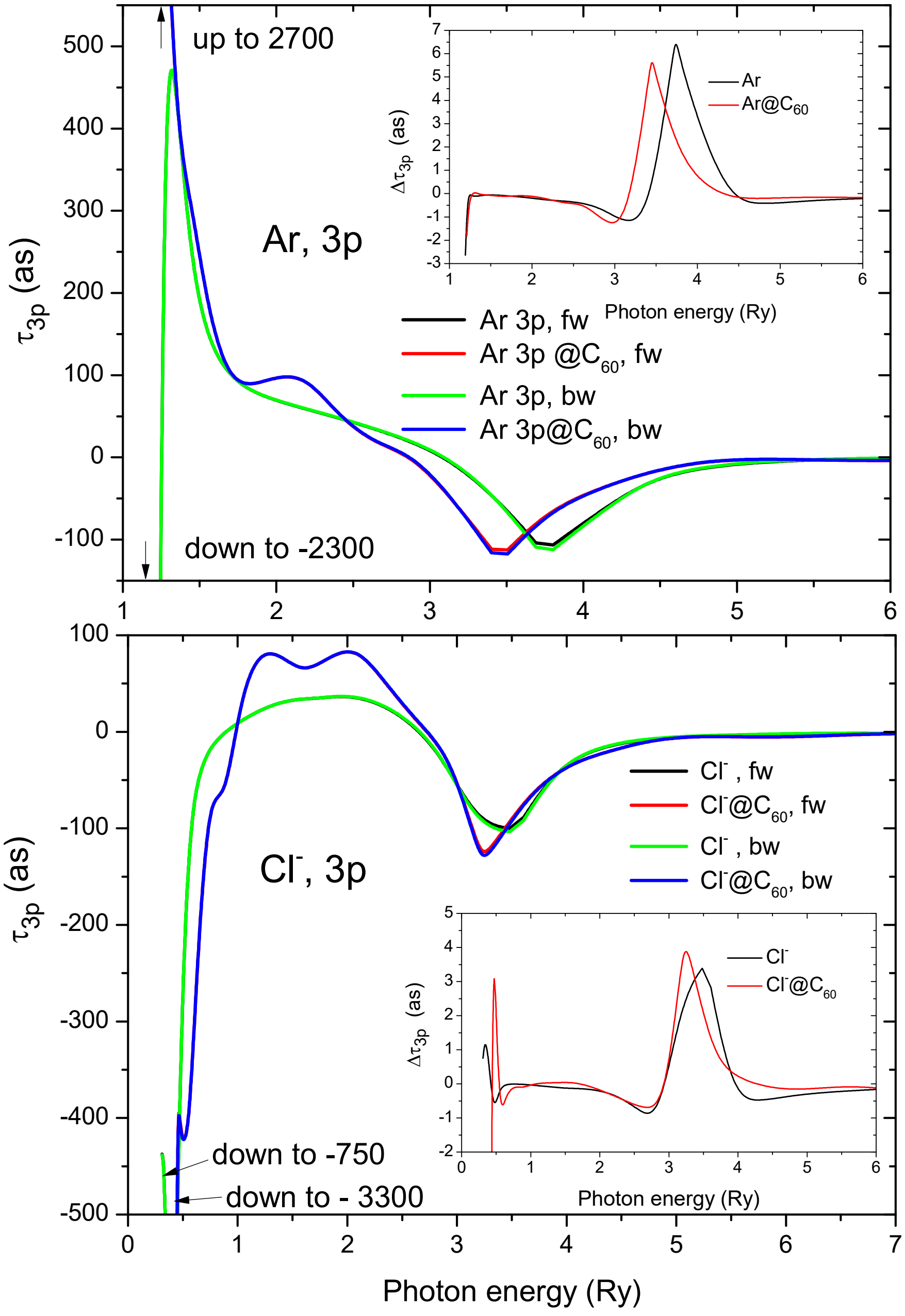

Fig. 1. Time delays of photoelectrons from 3p subshell of Ar, $Cl^-$ and endohedrals $Ar@C_{60}$ and $Cl^-@C_{60}$, emitted forward (fw) and backward (bw), and their differences $\Delta\tau_{3p}(\varepsilon) \equiv \tau_{3p}(\varepsilon,0) - \tau_{3p}(\varepsilon,\pi)$ (insets), determined by non-dipole contribution.

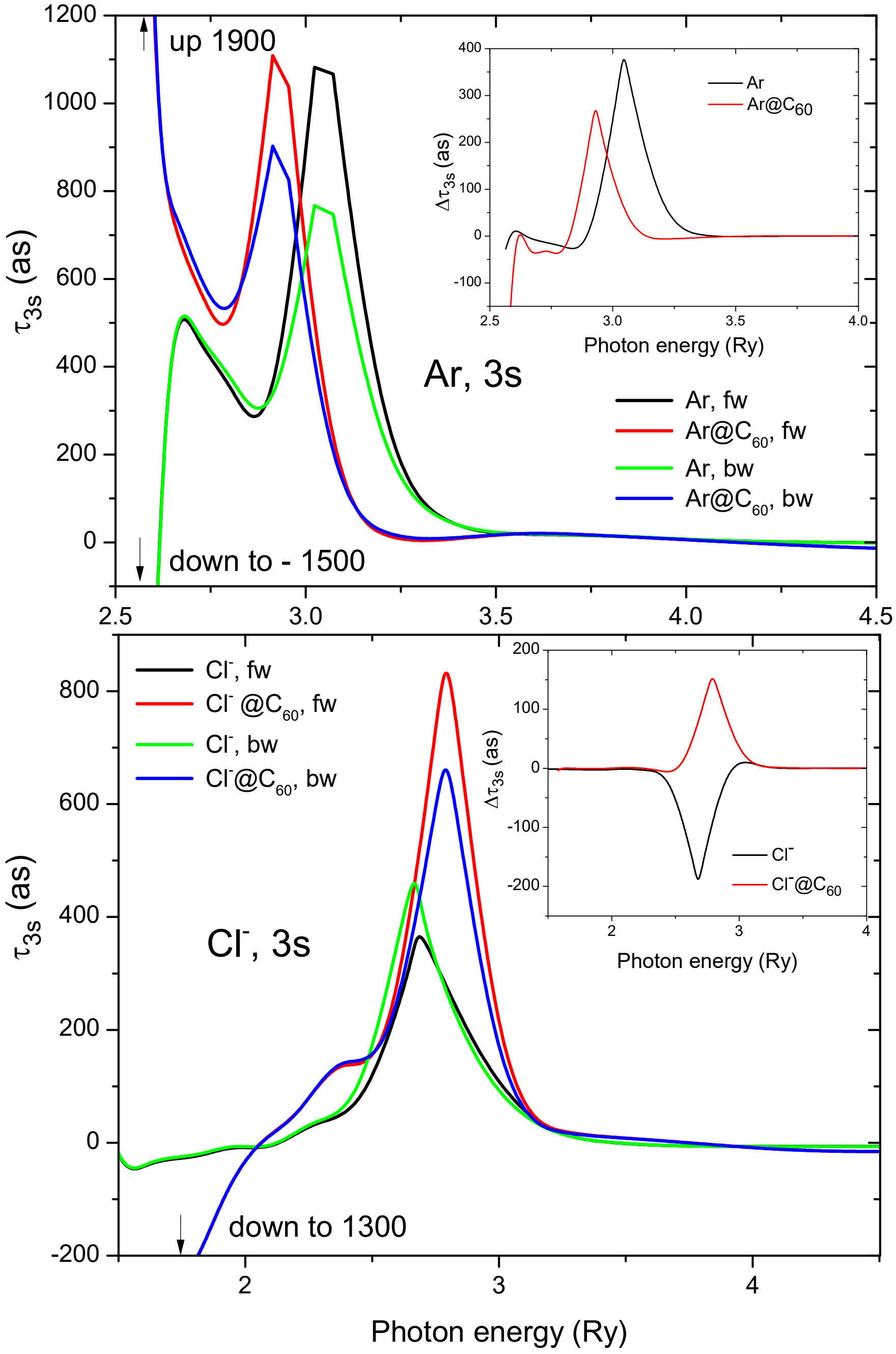

Fig. 2. Time delays of photoelectrons from 3s subshell of Ar, $Cl^-$ and endohedrals $Ar@C_{60}$ and $Cl^-@C_{60}$, emitted forward (fw) and backward (bw) and their differences $\Delta\tau_{3s}(\varepsilon) \equiv \tau_{3s}(\varepsilon,0) - \tau_{3s}(\varepsilon,\pi)$ (insets), determined by non-dipole contribution.

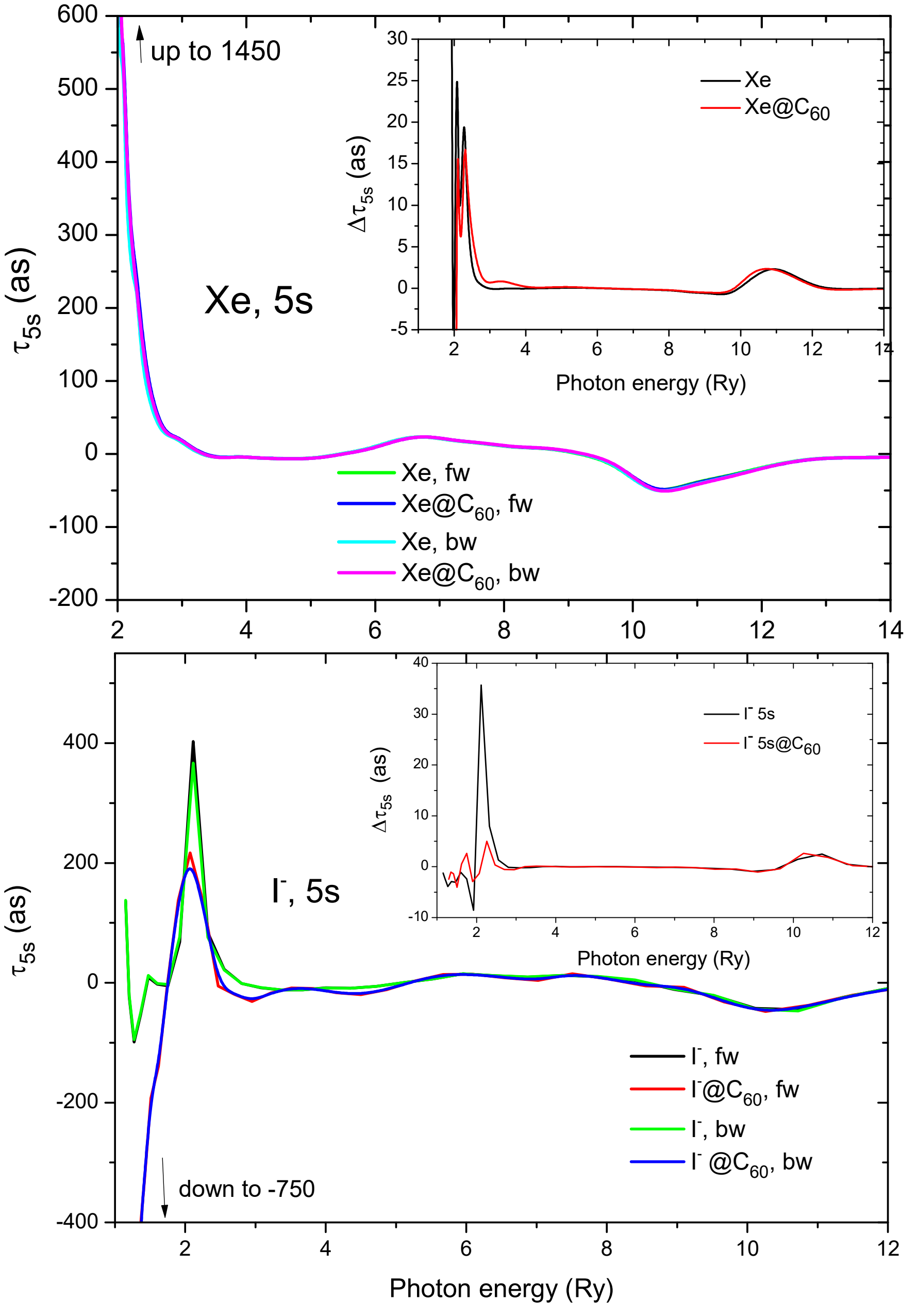

Fig. 3. Time delays of photoelectrons from 5s subshell of Xe, $I^-$ and endohedrals Xe@$C_{60}$ and $I^-$@$C_{60}$, emitted forward (fw) and backward (bw) and their differences $\Delta\tau_{5s}(\varepsilon) \equiv \tau_{5s}(\varepsilon,0) - \tau_{5s}(\varepsilon,\pi)$ (insets), determined by non-dipole contributions.

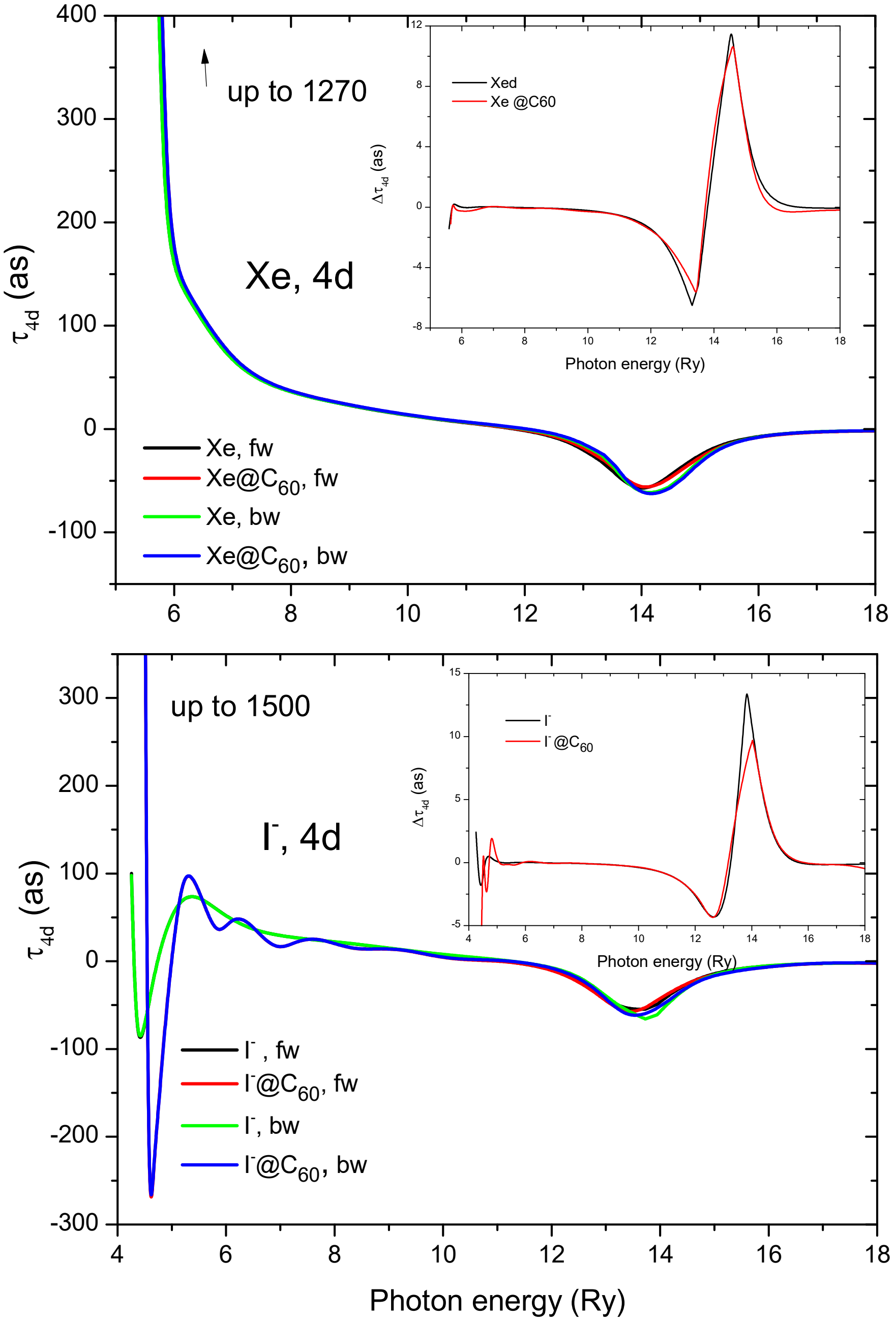

Fig. 4. Time delays of photoelectrons from 4d subshell of Xe, $I^-$ and endohedrals $Xe@C_{60}$ and $I^-@C_{60}$, emitted forward (fw) and backward (bw) and their differences $\Delta\tau_{4d}(\varepsilon) \equiv \tau_{4d}(\varepsilon,0) - \tau_{4d}(\varepsilon,\pi)$ (insets), determined by non-dipole contributions.